\documentclass[letter,twocolumn]{jpsj3}
\usepackage{txfonts}
\usepackage{bm}
\usepackage{color}
\title{$^{125}$Te-NMR Study on a Single Crystal of Heavy Fermion Superconductor UTe$_2$}

\author{Yo Tokunaga$^1$\thanks{E-mail: tokunaga.yo@jaea.go.jp}, Hironori Sakai$^1$, Shinsaku Kambe$^1$, Taisuke Hattori$^1$, Nonoka Higa$^1$, Genki Nakamine$^2$,
Shunsaku Kitagawa$^2$, Kenji Ishida$^2$, Ai Nakamura$^3$, Yusei Shimizu$^3$, Yoshiya Homma$^3$, DeXin Li$^3$, Fuminori Honda$^3$, and Dai Aoki$^{3,4}$}

\inst{$^1$ASRC, Japan Atomic Energy Agency, Tokai-mura, Ibaraki 319-1195, Japan  \\
$^2$Department of Physics, Kyoto University, Kyoto 606-8502, Japan  \\
$^3$IMR, Tohoku University, Oarai, Ibaraki, 311-1313, Japan  \\
$^4$University Grenoble Alpes, CEA, IRIG-PHELIQS, F-38000 Grenoble, France} 

\abst{
We report \Te~NMR studies on a newly discovered heavy fermion superconductor UTe$_2$.
Using a single crystal, we have measured the \Te~NMR Knight shift $K$ and spin-lattice relaxation rate $1/T_1$ for fields along the three orthorhombic crystal axes.
The data confirm a moderate Ising anisotropy for both the static ($K$) and dynamical susceptibilities ($1/T_1$) in the paramagnetic state above about 20 K. Around 20 K, however, we have observed a  sudden loss of NMR spin-echo signal due to sudden enhancement of the NMR spin-spin relaxation rate $1/T_2$, when the field is applied along the easy axis of magnetization (=$a$ axis).
This behavior suggests the development of longitudinal magnetic fluctuations along the $a$ axis at very low frequencies below 20 K. 
}

\newcommand{\UTe}{UTe$_2$}
\newcommand{\Te}{$^{125}$Te}

\begin{document}
\maketitle

Up to now, uranium-based compounds UGe$_2$ \cite{UGe2SC}, URhGe \cite{AokiNature} and UCoGe \cite{UCoGe} are the only fully established examples of ferromagnetic (FM) superconductors in which uniform superconductivity (SC) exists deep inside the FM state.
A characteristic feature of these FM superconductors is that they exhibit very large upper critical fields, by far exceeding the ordinary Pauli paramagnetic limit ($H_{\rm p}/T_{\rm sc}$=1.86 T/K). This provides a strong indication of equal-spin (triplet) SC pairing.\cite{Sheikin,LevyScience,Hardy2005,LevyNP,Aoki2009}.
Spin fluctuations near a FM quantum critical point (QCP) have been suggested to create the binding force between equal-spin pairs in these systems,\cite{Fay,Valls} analogous to the mechanism of super--fluid pairing in $^{3}$He.\cite{Levin}
Unconventional re--entrant or enhanced SC (RSC) phenomena observed in URhGe and UCoGe highlight a close interplay between magnetism and SC in these systems.\cite{LevyScience,Hardy2005,LevyNP,AokiJPSJ81,AokiJPSJ83}
The RSC is driven by a field-dependent pairing mechanism, ascribed to modulation of the excitation spectrum of FM fluctuations by external field, as demonstrated by recent NMR and thermal transport measurements.\cite{THattori,THattori2,YTPRL,YTPRB,Kotegawa,Wu,Mineev2011,Tada,Tada2,Hattori,Mineev2015} 

Recently, Ran {\it et al} reported evidence for SC in the uranium-based heavy fermion material UTe$_2$, exhibiting the rather high transition temperature of 1.6 K.\cite{RanUTe2}
Their finding was soon after confirmed by a more recent publication.\cite{AokiUTe2} 
The ground state of UTe$_2$ is paramagnetic, not ferromagnetic. However, the compound still exhibits a very large and anisotropic $H_{\rm c2}$, exceeding the Pauli limit along the three principal axes, similar to the above-mentioned FM superconductors. 
A sharp increase of $H_{\rm c2}$ for field along the $b$-axis is reminiscent of the RSC in URhGe and UCoGe.
Another interesting feature observed in UTe$_2$ is a discrepancy in the entropy balance at $T_{\rm sc}$
between SC and normal states, implying hidden features at low temperatures.\cite{AokiUTe2} 
A large residual value of the Sommerfeld coefficient below $T_{\rm SC}$ is found to be nearly half of the normal phase value\cite{RanUTe2,AokiUTe2}.  
For this reason a non-unitary triplet state, where only half of the Fermi sea of a
given spin direction is gapped, has been proposed.\cite{RanUTe2}

\begin{figure}[tbp]
\vspace{2mm}
\begin{center}
\includegraphics[width=8.5cm]{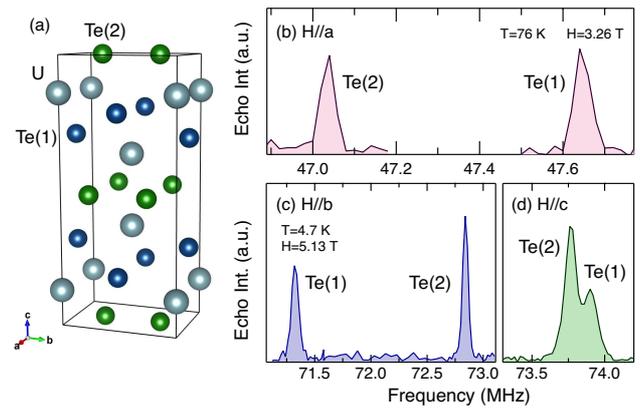}
\end{center}
 \vspace*{-10pt}
 \caption{(color online) (a) Schematic view of the body-centered orthorhombic crystal structure of \UTe. 
(b) Examples of \Te~NMR spectra measured with field along (b) the $a$-axis, (c) the $b$-axis, and (d) the $c$-axis.}
 \vspace*{-15pt}
  \label{Fig:Spectra}
\end{figure}

In UTe$_2$ the U-U distance of 3.78\AA\ is larger than the Hill limit ($\sim$3.5 \AA).\cite{Hill} 
A nearly localized nature for the U moment at high temperatures is indeed suggested from magnetic susceptibility data, which shows Curie-Weiss behavior above 150 K with an effective moment close to the   5f$^2$ or 5f$^3$ free ion value.\cite{IkedaUTe2,RanUTe2}
At low temperatures, the magnetic susceptibility increases rapidly with cooling for field along the easy $a$-axis, with a shoulder-like anomaly near 20 K.
In contrast, the susceptibility along the hard $b$-axis exhibits a broad maximum around 30 K.\cite{RanUTe2}  
For temperatures lower than 9 K and fields smaller than 3 T, the $a$-axis magnetization data have been found to be described by the Belitz-Kirkpatrick-Vojta (BKV)
theory of metallic FM quantum criticality.\cite{BKV} Thus, UTe$_2$ has been proposed to be a system close to a FM quantum critical point, dominated by strong magnetic fluctuations.\cite{RanUTe2,BKV}

In this paper we report a microscopic investigation of magnetic anisotropy and fluctuations in \UTe.
Orientation-dependent $^{125}$Te NMR measurements on a single crystal have confirmed
 a moderate Ising anisotropy for both the static and dynamical susceptibilities above about 20 K. Below 20 K, however, the NMR detected strong longitudinal magnetic fluctuations developed along the easy-axis of magnetization (=$a$-axis) at very low frequency when the field is applied along the $a$ axis, suggesting the occurrence of a  characteristic crossover temperature of magnetic fluctuations in the paramagnetic state.
 

\begin{figure}[tbp]
\vspace{5mm}
\begin{center}
\includegraphics[width=8.5cm]{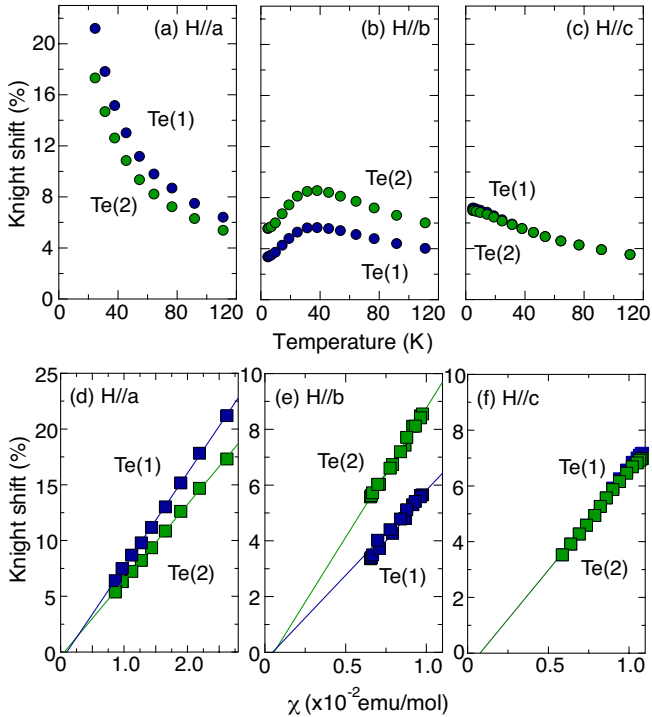}
\end{center}
\vspace*{-15pt}
\caption{(color online) Upper panels (a)-(c) show the temperature dependence of the NMR (Knight) shift for fields applied along the three crystalline axes [(a) $H$= $3.26$\,T, (b) and (c) $5.13$\,T].
In the lower panels (d)-(f), Knight shifts are plotted against the bulk susceptibility $\chi$ with temperature as an implicit parameter. 
The slopes of the lines yields the transferred HF coupling constants $A_{\rm hf}$ given in TABLE I. 
}
\vspace*{-15pt}
\label{Fig:K}
\end{figure}


Single crystals of \UTe~were grown using the chemical vapor transport method with iodine as transport agent.\cite{AokiUTe2}. The resistivity data of our crystals are consistent with previous results, showing a clear anomaly for the superconducting transition at $T_{\rm SC}=$1.65 K.\cite{RanUTe2,AokiUTe2} 
The $^{125}$Te NMR measurements were carried out on a single crystal (2$\times$2$\times$1 mm size) using a superconducting magnet and a phase--coherent, pulsed spectrometer.   The $^{125}$Te nuclei have a natural abundance of 7.0\% and the nuclear gyromagnetic ratio $\gamma_{\rm N}/2\pi$=13.454 MHz/T ($I$=$1/2$).

The NMR spectra were measured by recording integrated spin-echo intensities while sweeping frequency at a fixed external field.  The temperature dependence of the NMR (Knight) shift was derived from the peak position of the NMR spectrum with field applied along the three respective
orthorhombic crystal axes, where Cu NMR signals from metallic copper were used as markers for field calibrations.  
The spin-lattice relaxation rate $1/T_1$ was also measured at several different temperatures with fields along the three axes.
The measured nuclear magnetization recovery was found to fit a simple exponential for these $I=1/2$ nuclei.  Values of $1/T_2$ were determined by fitting the $\tau$ dependence of the spin-echo intensity to an exponential function, $M(2\tau)\propto \exp(-2\tau/T_2)$, as shown below. 

Figure 1(a) shows the crystal structure of UTe$_2$, which is the body-centered orthorhombic structure with space group $Immm$ (No.\,71, $D_{2h}^{25}$). There are four formula units per unit cell, and the 4U atoms occupy 4$i$ site. The 8Te atoms occur at two different sites: 4$j$, and 4$h$ sites with point symmetries $mm2$, and $m2m$, respectively. 
We denote these respective sites as Te(1) and Te(2), as seen in Fig.\,1(a).

In Fig.\,1(b)-(d) we display examples of $^{125}$Te NMR spectra obtained using our single crystal sample with the field applied in turn along the three respective
orthorhombic crystal axes, $\alpha =a, b$ and $c$.
The spectrum consists of two distinct peaks arising from the Te(1) and Te(2) sites for $H\|a$ and $b$. For $H\|c$, on the other hand, the two peaks are nearly overlapped.
Since $^{125}$Te nuclei are $I=1/2$, there is no quadrupolar splitting nor broadening.
We note that the site assignment presented here is not unambiguous; it is based on the anisotropy of the hyperfine coupling, which should reflect the local arrangement of neighboring U atoms of each Te sites (see the supplemental materials for details).
Further experiments will be required for exact identifications, but the assignment is not crucial for the following discussion in this paper.

In Figs.\,2 (a)-(c), we show the temperature dependence of the Knight shifts ($K_{\alpha}$) measured for fields applied along the three crystalline axes.
The $K_{\alpha}$ are nearly isotropic at high temperatures.
With decreasing temperature, $K_{a}$ exhibits a monotonic and rapid increase, while $K_{b}$ shows a broad maximum around 30 K.  
$K_{c}$ is in between, showing a monotonic and gradual increase.
These behaviors have essentially the same origin as bulk magnetic susceptibility, $\chi_{\alpha}(T)$.
For $H\|a$ no Knight shift values were obtained below 20 K, since we observed a sudden loss of the NMR spin-echo signal in that temperature range.
We will discuss this characteristic behavior later.

 \begin{table}[b]
\caption{ Hyperfine coupling constants $A_{\rm hf}^\alpha$ evaluated from $K_{\alpha}$ vs $\chi_{\alpha}$ plots in Fig.\,2(d)-(f).   }
  \begingroup
\begin{tabular}{@{\hspace{0.25cm}}c@{\hspace{0.5cm}}c@{\hspace{0.25cm}}c@{\hspace{0.5cm}}c@{\hspace{0.25cm}}c@{\hspace{0.5cm}}c@{\hspace{0.25cm}}} \hline \hline
 &$H\|a$& &$H\|b$& &$H\|c$\\ \cline{2-2} \cline{4-4} \cline{6-6}
 & $A_{\rm hf}^{\rm a}$(kOe/$\mu _B$) & & $A_{\rm hf}^{\rm b}$(kOe/$\mu _B$) & & $A_{\rm hf}^{\rm c}$(kOe/$\mu _B$)\\ [3pt]
\hline
Te(1)   & 47.1  & & 34.1 & & 39\\ 
Te(2)   & 38.0& & 51.8 & & 39\\
  \hline \hline
\end{tabular}
\endgroup
\end{table}

In Figs.\,2(d)-(f), we plot $K_{\alpha}$ against $\chi_\alpha$ with temperature as an implicit variable.
The $K_{\alpha}$ in all directions maintains a good linear relation with $\chi(T)$, and thus the slope of the $K$ vs $\chi$ plots yields the hyperfine coupling constants $A_{\rm hf}$. The estimated values of $A_{\rm hf}$ are summarized in TABLE I.
$A_{\rm hf}$ is positive in all three directions and is roughly independent of the crystalline axis.
The $A_{\rm hf}$ values obtained here are definitely larger than values expected from the classical dipolar coupling mechanism, that is, $|A_{\rm dip}|\lesssim$ 1 kOe/$\mu _B$, confirming the dominance of the transferred hyperfine coupling mechanism.
The transferred hyperfine coupling in a metal arises from the on-site spin density transferred from magnetic ions through spin polarization of the conduction electrons. 

Next, we discuss the nature of magnetic fluctuations in UTe$_2$.
Figures 3(a)-(c) show the temperature dependence of $(1/T_1T)_\alpha$ for fields applied along all three crystal axis directions ($\alpha$= $a$, $b$ and $c$). 
$(1/T_1T)_{b,c}$ exhibit a strong temperature dependence in contrast with flat behavior for $(1/T_1T)_{a}$.
With decreasing temperature, only $(1/T_1T)_{b,c}$ increase rapidly and tend to saturate below 10 K.
As discussed below, since $1/T_1$ is determined by spin fluctuations perpendicular to the quantization axis of the nuclear spins, i.e. the fixed field axis, 
this characteristic anisotropy in $1/T_1T$ indicates the dominance of magnetic fluctuations along the $a$ axis. 

\begin{figure}[tbp]
\vspace{2mm}
\begin{center}
 \includegraphics[width=8.5cm]{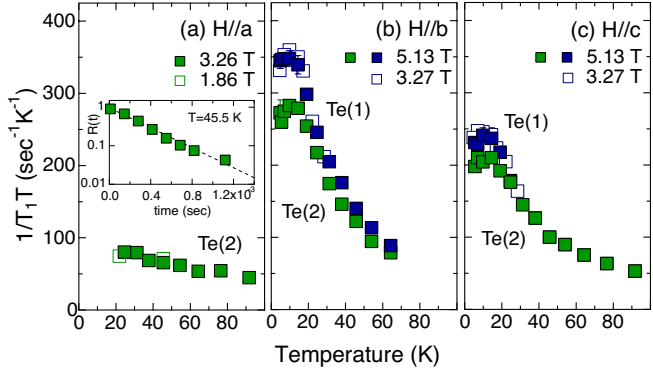}
 \end{center}
 \vspace*{-15pt}
 \caption{(color online) The temperature dependence of $(1/T_1T)_\alpha$ for fields applied along the three crystalline axes.
No obvious field-dependence of the $(1/T_1T)_\alpha$ has been observed in the field region of this study. 
 The inset shows the nuclear magnetization recovery against pulse duration time obtained at $T$=45.5 K and $H$=$3.26$\,T along the $a$ axis. }
\vspace*{-15pt}
\label{Fig:T1}
\end{figure}


In general, $1/T_1T$ measured in a field along the $\alpha$ direction is associated with the imaginary part of the
dynamic susceptibility $\chi^{''}_{\beta,\gamma}({\bm{q}}, \omega_{\rm n})$ along the $\beta$ and $\gamma$ directions perpendicular
to $\alpha$, \cite{Moriya,Ihara}
\begin{equation}
\label{eqinvT1T}
\left(\frac{1}{T_{1}T}\right)_{\alpha}=\frac{\gamma_{\rm n}^{2}k_{\rm B}}{2}\sum_{{\bm q}}\left[|A_{\rm hf}^{\beta}|^2\frac{\chi^{''}_{\beta}({\bm q},\omega_{\rm n})}{\omega_{\rm n}}+|A_{\rm hf}^{\gamma}|^2\frac{\chi^{''}_{\gamma}({\bm q},\omega_{\rm n})}{\omega_{\rm n}}\right],
\end{equation}
where $\omega_n$ is the NMR resonance frequency and $A_{\rm hf}$ is the hyperfine coupling constant for the $^{125}$Te nucleus.
In this way we have evaluated the directional dynamic susceptibility components for each orthorhombic crystal axis,
\begin{equation}
 R_ \alpha=\sum_{{\bm q}}|A^{\alpha}_{\rm hf}|^2\frac{\chi^{''}_{\alpha}({\bm q},\omega_{\rm n})}{\omega_{\rm n}}
 \end{equation} 
using the relations $(1/T_1T)_a=R_b+R_c$, $(1/T_1T)_b=R_a+R_c$ and $(1/T_1T)_c=R_a+R_b$.  The results are shown in Fig.\,4.
Since we have no $(1/T_1T)_a$ data below 20 K (due to the disappearance of the NMR signal), we can not estimate $R_{\alpha}$ in that temperature region.
\begin{figure}[tbp]
\vspace{2mm}
\begin{center}
\includegraphics[width=8.0cm]{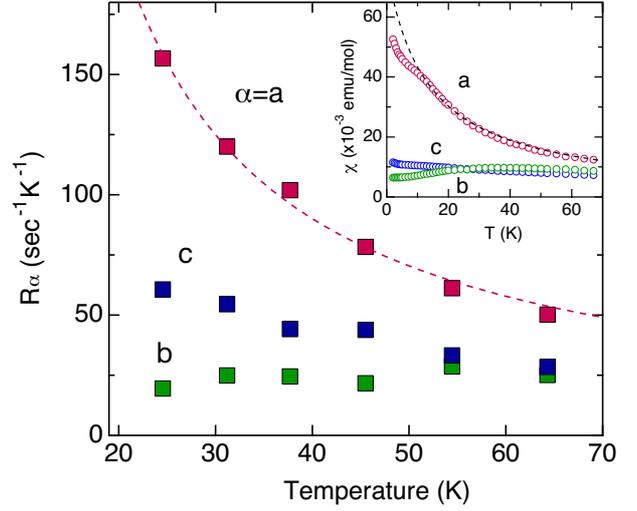}
 \end{center}
 \vspace*{-15pt}
 \caption{(color online) Orientation-resolved dynamic susceptibility $R_\alpha$ (see text) along the three crystalline axis directions in UTe$_2$.
The Curie-Weiss fit to the data for $R_{a}$ above 20 K (the dashed line) gives a small and positive value for the Weiss temperatures, i.e., $\theta_{\rm CW}\simeq4$ K.
The inset shows the temperature dependence of magnetic susceptibility along the three crystalline axes at $H=0.5$ T.
}
\vspace*{-15pt}
\label{Fig:Ra}
\end{figure}

\begin{figure}[tbp]
\vspace{2mm}
\begin{center}
 \includegraphics[width=7.5cm]{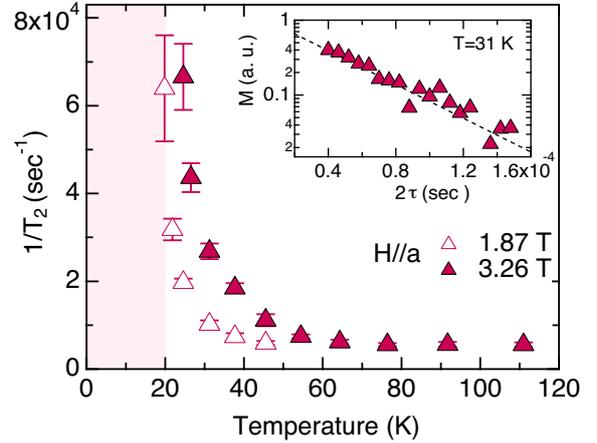}
 \end{center}
 \vspace*{-15pt}
 \caption{(color online) The temperature dependence of $1/T_2$ for fields applied along the the $a$- axis.
The inset shows a spin-echo decay curve obtained $T$=31 K and $H$=3.26 T($\|a$).}
\vspace*{-15pt}
\label{Fig:T2}
\end{figure}


Figure\,4 demonstrates Ising-type anisotropy for the dynamical spin susceptibilities, i.e., $R_{a}\gg R_{c}> R_{b}$ above 20 K.
This is, however, moderate if we compare it with the case of UCoGe.\cite{Ihara}
The static  ($K_{\alpha}$) and dynamical spin susceptibilities ($R_{\alpha}$) often possess contrasting anisotropies in $f$-electron systems.\cite{Kambe,Sakai1,Baek,Sakai2}  In the present case, however, we found nearly the same anisotropy for them. 
The bulk spin susceptibility ($\chi_{\alpha}$) shows a similar anisotropy and temperature dependence, as seen in the inset to Fig.\,4.
The scaling between the static and dynamical spin susceptibilities is anticipated for three dimensional (3D) FM fluctuations on the basis of self
consistent renormalization (SCR) theory.\cite{Moriya2}
Interestingly, both $(1/T_1T)_{b,c}$ and  $\chi_a(T)$ exhibit a deviation from the Curie-Weiss-like behavior around 20 K. 
At lower temperatures, however, $(1/T_1T)_{b,c}$ shows only a broad maximum, even though $\chi_a$ exhibits another strong upturn.
Instead, we have observed a strong enhancement of $1/T_2$ in that temperature region, as described below.

In Fig. 5 we show the temperature dependence of $1/T_2$ measured for fields applied along the $a$ axis. 
At higher temperatures, $1/T_2$ is nearly temperature independent. However,  $1/T_2$  starts to increase sharply upon cooling below about 40 K, and 
finally diverges around 20 K.  
The  divergence of $1/T_2$ has been observed only when the field is applied along the easy $a$-axis (not along the $b$ and $c$ axes).
Such an anisotropic $1/T_2$ behavior would be attributed to the electronic contribution,  which in general consists of two terms, $(1/T_2)^{\rm
el}=(1/T_2)^{\rm el}_{\|}+(1/T_2)^{\rm el}_{\bot}$ with
$(1/T_2)^{\rm el}_{\|}\propto G_{\|}(0)$ and $(1/T_2)^{\rm%
el}_{\bot}\propto G_{\bot}(\omega_{\rm NMR})$, where $\omega_{\rm
NMR}$ is the NMR frequency and
$G_{\alpha}(\omega)=\int_{-\infty}^{\infty}\left<h_{\alpha}(t)
h_{\alpha}(0) \right>\exp(i\omega t)dt$ is the spectral density of
the fluctuating hyperfine field, $h_{\alpha}(t)$.
Thus, $(1/T_2)^{\rm el}_{\|}$ is driven by the longitudinal
component of magnetic fluctuations around zero
frequency, while $(1/T_2)^{\rm el}_{\bot}$ is driven by the transverse ($\bot$) components of the fluctuations at the NMR
frequency, still a very low value ($\sim$0.05 GHz). The latter fluctuations also generate
the nuclear spin-lattice relaxation process.

In the present case,  neither $(1/T_1T)_{b}$ nor $(1/T_1T)_{c}$ exhibits the divergence around 20 K.
Thus, the transverse ($\bot$) components of the fluctuations at the NMR frequency is not the source of the 
divergence of $1/T_2$. Namely, the divergence is attributed to the longitudinal ($a$-axis) component of magnetic fluctuations around zero frequency.
In general, the development of such a low frequency mode of fluctuations implies the onset of static order along the $a$-axis.
However, neither specific heat nor other bulk measurements have detected any signature of a phase transition around 20 K.\cite{RanUTe2,AokiUTe2,IkedaUTe2}
Thus, we suppose that 20 K is a kind of crossover temperature for magnetic fluctuations along the $a$ axis.
Another possibility would be that the fluctuations are field-induced, i.e., appearing only with a field component along the $a$-axis. Indeed, there is always the $a$-axis field component during the measurements of $\chi_a$ and $1/T_2$, but not that of $(1/T_1T)_{b,c}$.
At the present time, however, only a small field dependence has been observed in $1/T_2$ (Fig.\,5).
Further careful analysis of the field dependence of $1/T_2$ is called for, perhaps including the variation with applied field and NMR frequeny.

An interesting question would be whether these strong longitudinal fluctuations survive at very low temperatures where the SC appears. 
So far, we can not answer this question, since we did not observe an NMR signal for $H\|a$ below 20 K ( which in and of itself implies the existence of strong fluctuations).
If the fluctuations survive, one might expect a close interaction with the SC. Indeed, a similar divergence of $1/T_2$ has been observed in the field region where the RSC occurs in URhGe.\cite{YTPRL,Kotegawa,YTPRB}
The large residual value of the Sommerfeld coefficient in the SC state of UTe$_2$\cite{RanUTe2,AokiUTe2} 
might also be associated with strong fluctuations at low temperatures.
Further NMR studies at lower temperatures would help to clarify these points.

In summary, we report \Te~NMR performed on a single crystal of UTe$_2$.
The Knight shift and $1/T_1T$ data confirm a moderate Ising anisotropy for both the static and dynamical susceptibilities above 20 K.
However, a sudden increase of $1/T_2$  toward 20 K for $H\|a$ suggests the development of strong longitudinal magnetic fluctuations along the $a$-axis at very low frequency.
We suppose that UTe$_2$ is close to a FM instability, and there is a characteristic crossover temperature for magnetic fluctuations around 20 K.

\section*{Acknowledgment}
We would like to thank R. E. Walstedt and Y. Yanase for valuable discussions.
A part of this work was supported by JSPS KAKENHI Grant Number JP15KK0174, JP15H05745, JP15H05884, JP15H05882, JP15K21732, JP16KK0106, JP16H04006, 19K03726, JP19H00646, ICC-IMR, and the REIMEI Research Program of JAEA.

\newpage

{\bf \noindent Supplementary Materials for \\$^{125}$Te-NMR Study on a Single Crystal of Heavy Fermion Superconductor UTe$_2$}

\vspace{0.25cm}

\section{The site assignment of $^{125}$Te NMR lines}
The crystal structure of UTe$_2$ is the body-centered orthorhombic structure with space group $Immm$ (No.\,71, $D_{2h}^{25}$).
In this structure there are two non--equivalent Te sites, Te(1)-4$j$ and Te(2)-4$h$, providing two NMR lines at different frequencies in Fig. 1(b)-(d). 
In Fig.\,6, we show the angular dependence of the NMR frequencies measured by means of a two-axis
sample rotator installed in the experimental cryostat.
Both of the NMR lines change frequency with sample rotation, (a), from the $a$ to the $c$ axes, and (b), from the $c$ to the $b$ axes. 
These results confirm that the NMR signals arise from a single crystal of UTe$_2$.
\begin{figure}[htb]
\vspace{-3mm}
\begin{center}
\includegraphics[width=8.5cm]{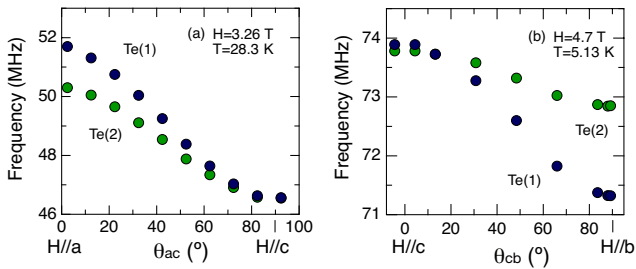}
\end{center}
 \vspace*{-6pt}
 \caption{The field-angular dependence of the NMR frequencies, which were measured by means of a two-axis
sample rotator installed in the experimental cryostat. Thus, the field is rotated from the $a$ to the $c$ axes and from the $c$ to the $b$ axes for the (a) and (b) sites, respectively.
}
\vspace*{-15pt}
  \label{Fig:Spectra}
\end{figure}

\begin{figure}[htb]
\vspace{-10mm}
\begin{center}
\includegraphics[width=9.0cm]{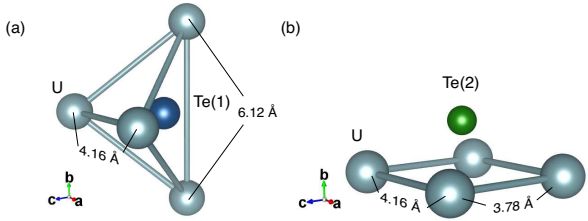}
\end{center}
 \vspace*{-12pt}
 \caption{Schematic views of the local arrangement of U atoms around  the (a), Te(1), and (b), Te(2) sites.}
 \vspace*{-15pt}
  \label{Fig:Spectra}
\end{figure}
The site assignment of the NMR lines presented in the main article is based on the anisotropy of the hyperfine coupling constants $A_{\rm hf}^\alpha$, which should reflect the local arrangement of neighboring U atoms around each Te site.
The estimated values of $A_{\rm hf}^\alpha$ for each of the two NMR lines are summarized in Table I of the main article.
The configurations of U atoms surrounding the two Te sites are presented in Fig.\,7. 

 As seen in Table I,  we have found that one of the NMR lines possesses a nearly isotropic value of $A_{\rm hf}^\alpha$  in the $ac$ plane with a larger value along the $b$-axis, i.e., $A_{\rm hf}^b > A_{\rm hf}^a \simeq A_{\rm hf}^c$.
Such an axial-type anisotropy is expected with the Te(2) sites, where the four nearest neighbor (n.n.) U atoms form a square-like lattice in the $ac$ plane, and the Te(2) atom lies on the top of that [Fig. 7(b)]. For Te(1) sites, on the other hand, the first and second n.n.\,U atoms form a distorted tetrahedral structure, which is stretched along the $b$-axis, as seen in Fig.\,7(a).
This structure provides anisotropy consistent with the other NMR line, that is, $A_{\rm hf}^a > A_{\rm hf}^c > A_{\rm hf}^b$.

\section{Comparison with NMR data from powdered crystals.}
 In Ref.[1], Te NMR data obtained in powder samples of UTe$_2$ are presented.
 A single, narrow NMR spectrum has been observed below 1.8 K.
 However, the estimated value of the Knight shift from the spectrum is only about 0.01\%, which is more than two orders of magnitude smaller than the values reported here for a single crystal at low temperatures ( i.e., $\sim20$\% for $H\|a$, and $\sim3$\% for $H\|b$) [see also Fig.2(a)-(c)]. 
This implies that the observed NMR spectrum reported for powdered crystals does not come from the main UTe$_2$ phase. 
 
In studies on a single crystal of UTe$_2$ we have observed a large and highly anisotropic $^{125}$Te Knight shift at low temperatures. It follows that NMR spectra observed with powdered crystals will be extremely broad if the grains of the powder are fixed so that the orientation of the magnetic field ($H$) is distributed randomly.  On the other hand, if the grains were not fixed, each grain could be aligned with the $a$-axis parallel to $H$ at low temperatures, owing to a large magnetic anisotropy of UTe$_2$.  In that case, however, the detection of the NMR signal might be even more difficult, since $T_2$ becomes extremely short below 20 K for $H\|a$, as was observed in the present NMR study.

\vspace{1cm}
\noindent
[1] S. Ran, C. Eckberg, Q.-P. Ding, Y. Furukawa, T. Metz, S. R. Saha, I.-L.Liu, M. Zic, H. Kim, J. Paglione, and N. P. Butch: arXiv:1811.11808.

\end{document}